# Simultaneous generation of two THz waves with bulk LiNbO$_3$ and four THz waves with PPLN by coupled optical parametric generation*

Zhongyang Li(李忠洋)[1†], Bin Yuan(袁斌)[1], Yongjun Li(李永军)[1], Lian Tan(谭联)[1], Pibin Bing(邴丕彬)[1], Hongtao Zhang(张红涛)[1] and Jianquan Yao(姚建铨)[2]

[1] *College of Electric Power, North China University of Water Resources and Electric Power, Zhengzhou 450045, China*

[2] *College of Precision Instrument and Opto-electronics Engineering, Institute of Laser and Opto-electronics, Tianjin University, Tianjin, 300072, China*

We present a theoretical research concerning simultaneous generation of two terahertz (THz) waves with bulk LiNbO$_3$ and four THz waves with periodically poled LiNbO$_3$ (PPLN) by coupled optical parametric generation (COPG). First, we investigate collinear phase matching of COPG generating two orthogonally polarized THz waves with two types of phase matching of $o = e + e$ and $o = e + o$ with bulk LiNbO$_3$. The two orthogonally polarized THz waves are generated from stimulated polariton scattering (SPS) with $A_1$ and $E$ symmetric transverse optical (TO) modes in bulk LiNbO$_3$, respectively. Then, we find that perturbations of phase mismatch for $o = e + e$ and $o = e + o$ can be compensated by a same grating vector of PPLN. As a result, four THz waves are simultaneously generated with a PPLN crystal and a pump laser. We calculate third-order nonlinear optical coefficients of $o = e + o$ generating THz waves from $E$ symmetric TO modes. The intensities of four THz waves are calculated by solving coupled wave equations. The calculation results demonstrate that the COPG generating four THz waves have high photon conversion efficiencies.

## 1. Introduction

Terahertz (THz) technology is of increasing interest for a broad range of scientific research and industrial applications, such as time-domain spectroscopy [1], molecular dynamics [2], imaging [3], telecommunications [4], chemical sensing [5], and quality monitoring in manufacturing [6]. A THz wave source with a high-power output and a high quantum efficiency is essential for above applications. Optical parametric process has proven to be an efficient scheme to generate THz waves [7–10]. The generation of THz waves by optical parametric process results from stimulated polariton scattering (SPS), which consists of second-order and third-order nonlinear optical processes.

* Project supported by the National Natural Science Foundation of China (61735010 and 61601183); the Natural Science Foundation of Henan Province (162300410190); the Program for Innovative Talents (in Science and Technology) in University of Henan Province (18HASTIT023)
† Corresponding author. E-mail: thzwave@163.com

LiNbO$_3$ has been the most widely used crystal for THz wave generation via SPS [7-9]. Due to strong dispersion of bulk LiNbO$_3$ between the optical and THz regions, Cherenkov phase matching or noncollinear phase matching is utilized. For strong interactions among pump, Stokes and THz waves, quasi phase matching with periodically poled LiNbO$_3$ (PPLN) is employed. In the past, by utilizing the largest second-order nonlinear coefficient $d_{33}$ of LiNbO$_3$, THz waves were generated from $A_1$ symmetric infrared- and Raman-active transverse optical (TO) modes by SPS with type-0 noncollinear phase matching $e = e + e$ [7-9]. Recently, Akiba reported forward and backward THz wave generations using type-II collinear phase matching via difference frequency generation (DFG) with bulk LiNbO$_3$ [11]. The forward and backward THz waves were generated from $E$ symmetric TO modes.

LiNbO$_3$ has five $A_1$ symmetric infrared- and Raman-active TO modes polarized parallel to the $c$-axis with frequencies of 248 cm$^{-1}$, 274 cm$^{-1}$, 307 cm$^{-1}$, 628 cm$^{-1}$, and 692 cm$^{-1}$, and has eight $E$ symmetric infrared- and Raman-active TO modes polarized perpendicular to the $c$-axis with frequencies of 152 cm$^{-1}$, 236 cm$^{-1}$, 265 cm$^{-1}$, 322 cm$^{-1}$, 363 cm$^{-1}$, 431 cm$^{-1}$, 586 cm$^{-1}$, and 670 cm$^{-1}$ [12]. Both $A_1$ and $E$ symmetric TO modes can be used for THz wave generation via SPS. If the phase matching conditions of two SPS processes generating two THz waves from $A_1$ and $E$ symmetric TO modes with a pump laser can be simultaneously satisfied with bulk LiNbO$_3$, the two THz waves can be simultaneously generated with coupled optical parametric generation (COPG). Moreover, the perturbations of phase mismatch for the COPG can be compensated by a grating vector of PPLN. As a result, four THz waves can be simultaneously generated from the COPG with a PPLN crystal and a pump laser.

In this work, we theoretically study the simultaneous generation of two THz waves from two SPS processes with $A_1$ and $E$ symmetric TO modes with bulk LiNbO$_3$. By compensating perturbations of phase mismatch for the COPG, four THz waves are simultaneously generated from the COPG with a PPLN crystal and a single pump wave. We calculate third-order nonlinear optical coefficients of $o = e + o$ generating THz waves from $E$ symmetric TO modes. The intensities of four THz waves are calculated by solving coupled wave equations. In this work, the bulk LiNbO$_3$ is 5% MgO-doped congruent LiNbO$_3$ and the PPLN is 5% MgO-doped congruent PPLN. To our best knowledge, this is the first study on simultaneous generation of four THz waves from COPG processes. A pump laser simultaneously generates four THz waves and four Stokes waves with collinear phase matching, which indicates that the four THz waves and the four Stokes waves are phase-conjugate. Moreover, since the polarization directions of the four THz waves are orthogonal, the four THz waves can be used in imaging and spectral analysis.

## 2. Phase-matching characteristics

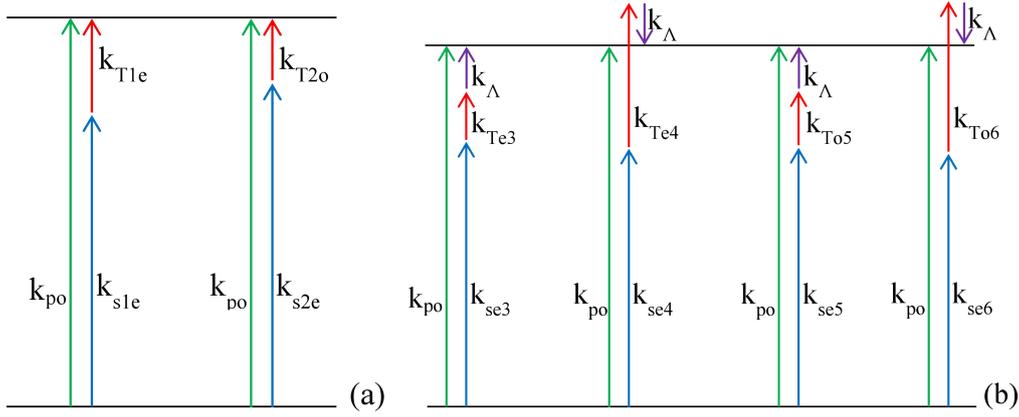

**Fig. 1.** Phase matching vectors of COPG. (a) Phase matching vectors of COPG generating two THz waves with bulk LiNbO$_3$. (b) Quasi phase matching vectors of COPG generating four THz waves with PPLN. The green, blue, red and purple arrows denote the wave vectors of the pump wave, Stokes wave, THz wave and grating vector respectively. The subscript "o" indicates ordinary wave and the subscript "e" indicates extraordinary wave.

As LiNbO$_3$ is negative uniaxial crystal, collinear phase matching cannot be satisfied in bulk LiNbO$_3$ using type-0 phase matching $e = e + e$. However, phase matching conditions of $o = e + e$ and $o = e + o$ can be satisfied in bulk LiNbO$_3$. The refractive index of an extraordinary wave is changed when $\theta$ is tuning in bulk LiNbO$_3$. The phase matching conditions of $o = e + e$ and $o = e + o$ can be simultaneously satisfied in bulk LiNbO$_3$ by choosing accurate $\theta$ and pump wavelength. Fig. 1 (a) shows phase matching vectors of COPG generating two THz waves with bulk LiNbO$_3$ with a same pump laser. For the COPG processes, the forward collinear phase matching conditions $o = e + e$ and $o = e + o$ are represented by

$$\left|\boldsymbol{k}_{\mathrm{po}}\right|-\left|\boldsymbol{k}_{\mathrm{s1e}}\right|=\left|\boldsymbol{k}_{\mathrm{T1e}}\right| \Rightarrow \frac{2\pi n_{\mathrm{p}}}{\lambda_{\mathrm{p}}}-\frac{2\pi n_{\mathrm{s1}}}{\lambda_{\mathrm{s1}}}=\frac{2\pi n_{\mathrm{T1}}}{\lambda_{\mathrm{T1}}}, \quad (1)$$

$$\left|\boldsymbol{k}_{\mathrm{po}}\right|-\left|\boldsymbol{k}_{\mathrm{s2e}}\right|=\left|\boldsymbol{k}_{\mathrm{T2o}}\right| \Rightarrow \frac{2\pi n_{\mathrm{p}}}{\lambda_{\mathrm{p}}}-\frac{2\pi n_{\mathrm{s2}}}{\lambda_{\mathrm{s2}}}=\frac{2\pi n_{\mathrm{T2}}}{\lambda_{\mathrm{T2}}}, \quad (2)$$

where $\boldsymbol{k}_{\mathrm{p}}$, $\boldsymbol{k}_{\mathrm{s}}$, $\boldsymbol{k}_{\mathrm{T}}$ are the wave vectors of the pump, Stokes and THz waves respectively, $\lambda_{\mathrm{p}}$, $\lambda_{\mathrm{s}}$, $\lambda_{\mathrm{T}}$ are the wavelengths of the pump, Stokes and THz waves respectively, $n_{\mathrm{p}}$, $n_{\mathrm{s}}$, $n_{\mathrm{T}}$ are the refractive indices of the pump, Stokes and THz waves respectively. The subscript "o" indicates ordinary wave and the subscript "e" indicates extraordinary wave. The subscript "1" indicates first set of Stokes and THz waves, and the subscript "2" indicates second set of Stokes and THz waves. The energy conservation conditions for the COPG processes generating two THz waves are represented by

$$\frac{1}{\lambda_p} - \frac{1}{\lambda_{s1}} - \frac{1}{\lambda_{T1}} = 0, \tag{3}$$

$$\frac{1}{\lambda_p} - \frac{1}{\lambda_{s2}} - \frac{1}{\lambda_{T2}} = 0, \tag{4}$$

Fig. 2 shows the phase matching conditions for $o = e + e$ and $o = e + o$ for $\theta$ rotation in bulk LiNbO$_3$ generating two THz waves with a pump wavelength of 1.55 μm. The dispersion data for LiNbO$_3$ in the calculation are taken from Ref. 13 in the optical region [13] and from Ref. 14 in the THz region [14]. From the figure we find that the two phase matching conditions of $o = e + e$ and $o = e + o$ generating two orthogonally polarized THz waves are simultaneously realized. Extraordinary THz waves in the range 0.1-4.28 THz from $o = e + e$ and ordinary THz waves in the range 0.1-2.84 THz from $o = e + o$ are generated by changing $\theta$ angle. At the same time, extraordinary Stokes waves in the range 1.55-1.585 μm from $o = e + e$ and in the range 1.55-1.573 μm from $o = e + o$ are generated by changing $\theta$ angle.

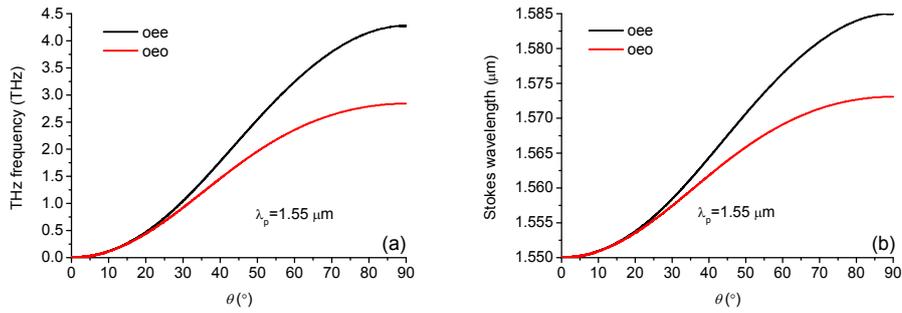

**Fig. 2.** Phase matching curves for $o = e + e$ and $o = e + o$ generating two THz waves in bulk LiNbO$_3$ with a pump wavelength of 1.55 μm. *oee* indicates that pump, Stokes and THz waves are ordinary, extraordinary and extraordinary waves, and *oeo* indicates that pump, Stokes and THz waves are ordinary, extraordinary and ordinary waves. (a) THz frequency versus $\theta$, (b) Stokes wavelength versus $\theta$.

In the processes of frequency tuning, Eqs. (1) and (2) cannot be satisfied at every THz frequency. The perturbations of phase mismatch for Eqs. (1) and (2) can be compensated by a grating vector $\boldsymbol{k}_\Lambda$ of PPLN. With $\boldsymbol{k}_\Lambda$, Eq. (1) is divided into two phase matching conditions Eqs. (5-6), and Eq. (2) is divided into two phase matching conditions Eqs. (7-8). Eqs. (5-8) are represented by

$$\left|\boldsymbol{k}_{po}\right| - \left|\boldsymbol{k}_{se3}\right| - \left|\boldsymbol{k}_{Te3}\right| = \left|\boldsymbol{k}_\Lambda\right| \Rightarrow \frac{2\pi n_p}{\lambda_p} - \frac{2\pi n_{s3}}{\lambda_{s3}} - \frac{2\pi n_{T3}}{\lambda_{T3}} = \frac{2\pi}{\Lambda}, \tag{5}$$

$$\left|\boldsymbol{k}_{po}\right| - \left|\boldsymbol{k}_{se4}\right| - \left|\boldsymbol{k}_{Te4}\right| = -\left|\boldsymbol{k}_\Lambda\right| \Rightarrow \frac{2\pi n_p}{\lambda_p} - \frac{2\pi n_{s4}}{\lambda_{s4}} - \frac{2\pi n_{T4}}{\lambda_{T4}} = -\frac{2\pi}{\Lambda}, \tag{6}$$

$$\left|\boldsymbol{k}_{po}\right|-\left|\boldsymbol{k}_{se5}\right|-\left|\boldsymbol{k}_{To5}\right|=\left|\boldsymbol{k}_\Lambda\right| \Rightarrow \frac{2\pi n_p}{\lambda_p}-\frac{2\pi n_{s5}}{\lambda_{s5}}-\frac{2\pi n_{T5}}{\lambda_{T5}}=\frac{2\pi}{\Lambda}, \tag{7}$$

$$\left|\boldsymbol{k}_{po}\right|-\left|\boldsymbol{k}_{se6}\right|-\left|\boldsymbol{k}_{To6}\right|=-\left|\boldsymbol{k}_\Lambda\right| \Rightarrow \frac{2\pi n_p}{\lambda_p}-\frac{2\pi n_{s6}}{\lambda_{s6}}-\frac{2\pi n_{T6}}{\lambda_{T6}}=-\frac{2\pi}{\Lambda}, \tag{8}$$

where $\Lambda$ is poling period of PPLN crystal. The subscript "3", "4", "5" and "6" indicate the third, fourth, fifth and sixth set of Stokes and THz waves.

The energy conservation conditions for the COPG processes generating four THz waves are represented by

$$\frac{1}{\lambda_p}-\frac{1}{\lambda_{sm}}-\frac{1}{\lambda_{Tm}}=0, \tag{9}$$

The subscript "$m$" = 3, 4, 5, 6 indicates the third, fourth, fifth and sixth set of Stokes and THz waves. Fig. 1 (b) shows quasi phase matching vectors of COPG generating four THz waves with PPLN with a same pump laser. Fig. 3 shows phase matching curves for quasi phase matching generating four THz waves in PPLN with a pump wavelength of 1.55 μm. From the figure we find that four quasi phase matching conditions which are described by Eqs. (5-8) generating four THz waves are simultaneously satisfied. With a fixed $\theta$, THz frequencies and Stokes wavelengths are tuning with $\Lambda$. The frequency differences of THz waves between two oee curves or that between two oeo curves are becoming smaller and smaller with the increasing $\Lambda$. The wavelength differences of Stokes waves between two oee curves or that between two oeo curves have the same trend. As $\theta$ changes from 24.6° to 39.8° and 45.6°, THz waves with frequencies from 0.4 to 2.5 THz are generated. Moreover, as $\theta$ changes from 0° to 90° which shows in Fig. 2, THz waves with frequencies from 0.1 to 4.28 THz can be generated. As shown in Fig. 2 and Fig. 3, wide tuning THz waves can be generated by selecting suitable $\Lambda$ and $\theta$.

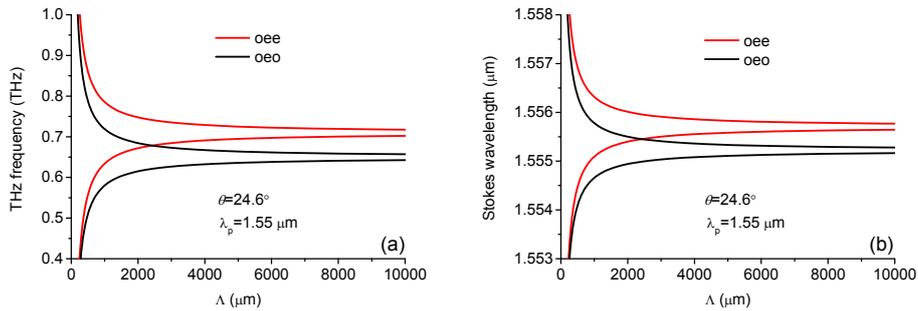

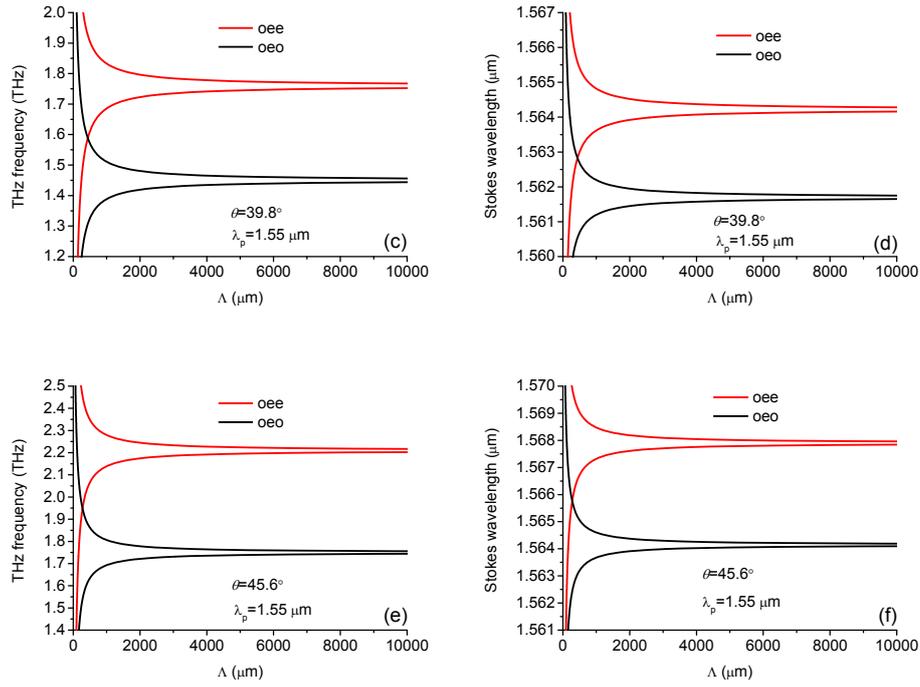

**Fig. 3.** Phase matching curves for quasi phase matching generating four THz waves in PPLN with a pump wavelength of 1.55 μm. oee indicated that the pump, Stokes and THz waves are ordinary, extraordinary and extraordinary waves, and oeo indicated that the pump, Stokes and THz waves are ordinary, extraordinary and ordinary waves. (a) THz frequency versus Λ with θ of 24.6°, (b) Stokes wavelength versus Λ with θ of 24.6°, (c) THz frequency versus Λ with θ of 39.8°, (d) Stokes wavelength versus Λ with θ of 39.8°, (e) THz frequency versus Λ with θ of 45.6°, (f) Stokes wavelength versus Λ with θ of 45.6°.

## 3. Nonlinear optical coefficients

Nonlinear optical coefficients are purely electronic when the frequencies of mixing waves involved in optical parametric process are far above the TO mode frequencies. However, when the frequencies of THz waves are near the TO mode frequencies, ionic nonlinearities as well as electronic nonlinearities are present. The analytical expression of nonlinear optical coefficient $d_m$ for COPG and THz wave absorption coefficient $\alpha_{Tm}$ for LiNbO$_3$ are written as follows [14]:

$$d_m = d_{em} + \sum_j \frac{S_j \omega_{TO_j}^2}{\omega_{TO_j}^2 - \omega_T^2} d_{Q_j}, \qquad (10)$$

$$\alpha_{Tm} = 2\frac{\omega_T}{c} \text{Im}(\varepsilon_\infty + \sum_j \frac{S_j \omega_{TO_j}^2}{\omega_{TO_j}^2 - \omega_T^2 - i\omega_T \Gamma_j})^{\frac{1}{2}}, \qquad (11)$$

where $d_{em}$ is the electronic second-order nonlinear coefficient, $d_Q$ is the ionic third-order nonlinear coefficient. $d_m$ is the bulk value of the nonlinear coefficient involving electronic and ionic contributions. For $m$=3 or 4, $d_m$ and $\alpha_{Tm}$ are for $A_1$ symmetric TO modes, and for $m$=5 or 6, $d_m$ and $\alpha_{Tm}$ are for $E$ symmetric TO modes.

$\omega_{TO}$, $S$ and $\Gamma$ denote eigenfrequency, oscillator strength and bandwidth of the TO modes respectively. The subscript "$j$" indicates the $j$th TO mode. $\omega_T$ is the angular frequency of THz wave. $\varepsilon_\infty$ is the high-frequency dielectric constant, and $c$ is the light velocity in vacuum. For $m=3$ or 4, $d_{em} = d_{22}\cos^2\theta\cos 3\varphi$, and for $m=5$ or 6, $d_{em} = d_{15}\sin\theta - d_{22}\cos\theta\sin 3\varphi$. We set $\varphi=0°$ for a maximum second-order nonlinear coefficient. $d_{Q_j} = [\dfrac{8\pi c^4 n_p (S_{igk}^j / Ld\Omega)}{S_j \hbar \omega_{TO_j} \omega_s^4 n_s (\overline{n}_T + 1)}]^{\frac{1}{2}}$ refers to the third-order Raman scattering, where $\overline{n}_T = (e^{\hbar\omega_T/kT} - 1)^{-1}$ is the Bose-Einstein distribution function, where $\hbar$ is Planck constant divided by $2\pi$, $k$ is Boltzman constant, $T$ is the temperature. $\omega_s$ is the angular frequency of a Stokes wave. $S_{igk}^j / Ld\Omega$ is the spontaneous-Raman scattering efficiency of the $A_1$ or $E$ symmetric TO modes, where $S_{igk}^j$ is the fraction of incident power which is scattered into a solid angle $d\Omega$ near a normal to the direction of the optical path of length $L$ [15]. $S_{igk}^j = S_{33}^j$ for $A_1$ symmetric TO modes, $S_{igk}^j = S_{42}^j$ for $E$ symmetric TO modes [16]. Based on reported parameter values in Ref. 16, we calculate values of $d_{Q_j}$ for $E$ symmetric TO modes at room temperature, as shown in Table 1. The vibration parameters for $A_1$ symmetric TO modes are reported in Ref. 14, and we show the parameters in Table 2.

**Table 1.** Vibration Parameters for $E$ symmetric TO modes at room temperature.

| $\omega_{TO_j}$ [12] (cm$^{-1}$) | $S_j$ [12] | $\Gamma_j$ [12] (cm$^{-1}$) | $\overline{n}_T + 1$ | $S_{42}^j / Ld\Omega$ [16] (10$^{-6}$cm$^{-1}$sr$^{-1}$) | $d_{Q_j}$ (pm/V) |
|---|---|---|---|---|---|
| 152 | 22 | 14 | 1.92 | 3.8 | 18.3 |
| 236 | 0.8 | 12 | 1.47 | 2.9 | 76.95 |
| 265 | 5.5 | 12 | 1.38 | 0.54 | 12.32 |
| 322 | 2.2 | 11 | 1.27 | 0.96 | 24.6 |
| 363 | 2.3 | 33 | 1.21 | 0.94 | 22.97 |
| 431 | 0.18 | 12 | 1.14 | 0.39 | 49.94 |
| 586 | 3.3 | 35 | 1.06 | 2.2 | 24.63 |

Table 2. Vibration Parameters for $A_1$ symmetric TO modes at room temperature [14].

| $\omega_{TO_j}$ (cm$^{-1}$) | $S_j$ | $\Gamma_j$ (cm$^{-1}$) | $\bar{n}_T + 1$ | $S_{33}^j/Ld\Omega$ (10$^{-6}$cm$^{-1}$sr$^{-1}$) | $d_{Q_j}$ (pm/V) |
|---|---|---|---|---|---|
| 248 | 16 | 21 | 1.43 | 16.0 | 40 |
| 274 | 1 | 14 | 1.34 | 4.0 | -76.66 |
| 307 | 0.16 | 25 | 1.25 | 0.95 | -93.32 |
| 628 | 2.55 | 34 | 1.05 | 10.2 | 60 |

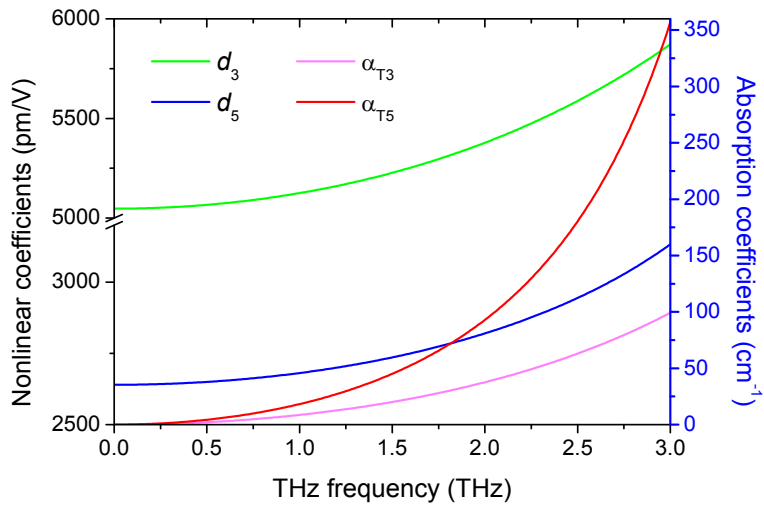

**Fig. 4.** Calculated nonlinear optical coefficients $d_3$ and $d_5$, and absorption coefficients $\alpha_{T3}$ and $\alpha_{T5}$ with $\lambda_p$ of 1.55 μm and $\theta$ of 24.6°.

Fig. 4 shows calculated nonlinear optical coefficients $d_3$ and $d_5$, and absorption coefficients $\alpha_{T3}$ and $\alpha_{T5}$ with $\lambda_p$ of 1.55 μm and $\theta$ of 24.6°. $d_{15}$ and $d_{22}$ are 4.6 and 2.76 pm/V with $\lambda_p$ of 1.55 μm [17]. From the figure we find that $\alpha_{T3}$ and $\alpha_{T5}$ smoothly and rapidly increase with frequencies. $\alpha_{T5}$ is larger than $\alpha_{T3}$, especially in high frequency region. At 3 THz $\alpha_{T3}$ and $\alpha_{T5}$ are 99 and 357 cm$^{-1}$ respectively. $d_3$ and $d_5$ smoothly and rapidly increase with frequencies. $d_3$ is obviously larger than $d_5$. At 3 THz $d_3$ and $d_5$ are 5872 and 3133 pm/V respectively. When frequencies approach the lowest TO mode frequency, the nonlinear optical coefficients and the absorption coefficients increase rapidly because polaritons induce giant ionic nonlinearities around the polariton resonance frequencies. As a result, $d_3$ and $d_5$ are much larger than second-order nonlinear coefficient $d_{em}$. The values of $d_3$ and $d_5$ at 3 THz for LiNbO$_3$ can compare with the nonlinear optical coefficient of 10000 pm/V for GaAs around the polariton resonance frequencies [18]. By comprehensive consideration of nonlinear optical coefficients and absorption coefficients, the intensities of THz wave generated from COPG with $A_1$ symmetric TO modes are larger than those with $E$ symmetric TO modes.

## 4. Intensities of four THz waves

In COPG, assume the pump, Stokes and THz waves are continuous plane waves with slowly varying envelopes co-propagating along the $+z$ direction. Under the slowly varying approximation, the coupled-wave equations for the COPG are represented by

$$\frac{\partial \vec{E}_p}{\partial z} = -i \sum_m \kappa_{pm} \vec{E}_{sm} \vec{E}_{Tm} e^{i\Delta k_m z}, \tag{12}$$

$$\frac{\partial \vec{E}_{sm}}{\partial z} = -i \kappa_{sm} \vec{E}_p \vec{E}_{Tm}^* e^{-i\Delta k_m z}, \tag{13}$$

$$\frac{\partial \vec{E}_{Tm}}{\partial z} = -\frac{\alpha_{Tm}}{2} \vec{E}_{Tm} - i \kappa_{Tm} \vec{E}_p \vec{E}_{sm}^* e^{-i\Delta k_m z}, \tag{14}$$

$$\kappa_p = \frac{\omega_p d_m}{c n_p}, \tag{15}$$

$$\kappa_{sm} = \frac{\omega_{sm} d_m}{c n_{sm}}, \tag{16}$$

$$\kappa_{Tm} = \frac{\omega_{Tm} d_m}{c n_{Tm}}, \tag{17}$$

where $\Delta k_m$ is the phase mismatch, $\omega_p$, $\omega_{sm}$ and $\omega_{Tm}$ are the angular frequencies of the pump, Stokes and THz waves respectively, $\kappa_p$, $\kappa_{sm}$ and $\kappa_{Tm}$ are the coupling coefficients of the pump, Stokes and THz waves respectively. With THz wave absorption and without phase mismatch, the coupled wave equations can be solved to give THz intensities, as shown in Fig. 5. In the calculations, the pump wave intensity $I_p$ is $10^4$ W/mm$^2$, and the initial Stokes wave intensities $I_{sm}$ are 100 W/mm$^2$. With $\lambda_p$ of 1.55 μm, $\Lambda$ of 1000 μm and $\theta$ of 24.6°, THz frequencies $\nu_{Tm}$ of the third, fourth, fifth and sixth THz waves are 0.63, 0.79, 0.58 and 0.72 THz respectively. From the figure we find that the intensities of the four THz waves increase rapidly with increasing crystal length to the maximum value and then decrease smoothly. The intensities of the third and fourth THz waves which are generated from $A_1$ symmetric TO modes are comparable to each other for $L > 10$ mm, and the same is true for the fifth and sixth THz waves which are generated from $E$ symmetric TO modes. The THz wave intensities generated from $A_1$ symmetric TO modes are much larger than those from $E$ symmetric TO modes because THz waves generated from $A_1$ symmetric TO modes have larger nonlinear coefficients and smaller absorption coefficients. The maximum intensities of the third, fourth, fifth and sixth THz waves are 6.2, 11.9, 0.3 and 0.44 W/mm$^2$ respectively, corresponding to the photon conversion efficiency of 18.9%, 29.3%, 1% and 1.2%, corresponding to the optimal crystal length of 76.9 mm, 74.1 mm, 62.6 mm and 59.7 mm. The difference length among the four optimal crystal lengths is small. We should choose the optimal crystal length at which the four THz waves are generated with a maximal total intensity. The total photon conversion efficiency from the pump wave to the four THz waves is 50.4%. The high photon conversion efficiency results from the following reasons. First, most of pump photons are converted to THz photons in the COPG. Second, absorption coefficients of the four THz waves are small since the frequencies of the four THz waves are below 1 THz. Third, in the calculations the quasi phase matching conditions of the COPG are satisfied.

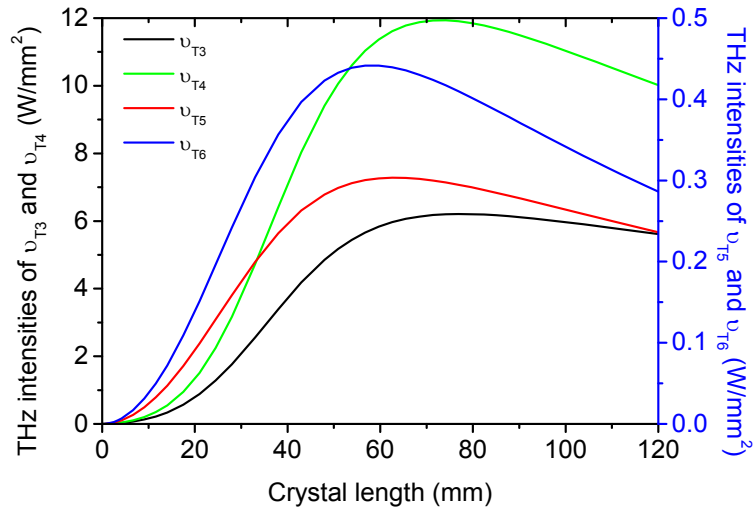

**Fig. 5.** THz wave intensities versus crystal length. Pump intensity $I_p$=10$^4$ W/mm$^2$, initial Stokes wave intensities $I_{sm}$=100 W/mm$^2$. $\lambda_p$ =1.55 μm, $\theta$ =24.6°, $\Lambda$=1000 μm. The frequencies of the $v_{T3}$, $v_{T4}$, $v_{T5}$ and $v_{T6}$ modes are 0.63, 0.79, 0.58 and 0.72 THz, respectively.

When the initial Stokes wave intensities $I_{s5}$ and $I_{s6}$ are enlarged, the intensities of the fifth and sixth THz waves are enhanced, as shown in Fig. 6. In the calculations $I_p$ =10$^4$ W/mm$^2$, $I_{s3}$ =100 W/mm$^2$, $I_{s4}$ =100 W/mm$^2$, $I_{s5}$=1600 W/mm$^2$, $I_{s6}$=2040 W/mm$^2$. From the figure we find that the intensities of the fifth and sixth THz waves are enhanced from those in Fig. 5. Compared with the intensities of the third and fourth THz waves in Fig. 5, the intensities of the third and fourth THz waves in Fig. 6 decrease. When the intensities of the fifth and sixth THz waves are enhanced, more pump wave photons consume in the COPG which generate the fifth and sixth THz waves. As a result, the intensities of the third and fourth THz waves decrease. By choosing the intensity parameters in Fig. 6, the maximum intensity of 6.6 W/mm$^2$ for the fourth wave is equal to that for the sixth THz wave, and the maximum intensity of 3.6 W/mm² for the third wave is equal to that for the fifth THz wave. As the polarization direction of the THz waves generated from $A_1$ symmetric TO modes and that of the THz waves generated from $E$ symmetric TO modes are orthogonal, we can choose THz waves with parallel or perpendicular polarization direction by a THz wire grid polarizer.

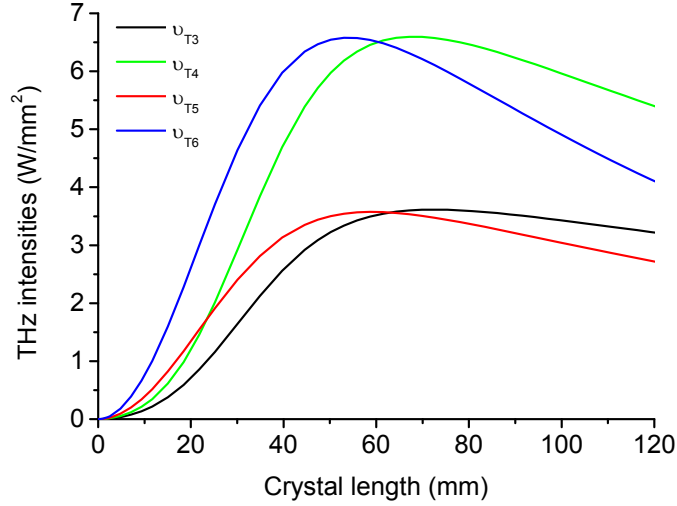

**Fig. 6.** THz wave intensities versus crystal length. $I_p=10^4$ W/mm², $I_{s3}=100$ W/mm², $I_{s4}=100$ W/mm², $I_{s5}=1600$ W/mm², $I_{s6}=2040$ W/mm². $\lambda_p=1.55$ μm, $\theta=24.6°$, $\Lambda=1000$ μm. The frequencies of the $v_{T3}$, $v_{T4}$, $v_{T5}$ and $v_{T6}$ modes are 0.63, 0.79, 0.58 and 0.72 THz, respectively.

The scheme proposed in this work has certain advantages. First of all, four THz waves are simultaneously generated only by a pump laser and a PPLN crystal. The pump laser, four Stokes waves and four THz waves are collinear, which means that the layout of experimental setups is simple. Second, the four THz waves which are generated from COPG with a same pump laser are phase-conjugate. Third, four THz waves with orthogonal polarization directions can be used in imaging and spectral analysis. However, the scheme has certain disadvantages. First, a PPLN crystal with a particular $\theta$ angle is difficult to manufacture. Second, the frequency tuning of four THz waves generated by a PPLN crystal is inconvenient. The frequency tuning of THz waves can be accomplished by changing poling period $\Lambda$. The poling period $\Lambda$ can be tuned by manufacturing a PPLN with a fanned-out structure.

## 5. Conclusion

COPG with two types of phase matching of $o = e + e$ and $o = e + o$ can generate two orthogonally polarized THz waves from $A_1$ and $E$ symmetric TO modes with bulk LiNbO₃. By compensating the perturbations of phase mismatch for $o = e + e$ and $o = e + o$ via a grating vector, COPG with quasi phase matching can generate four THz waves with PPLN. Compared with $E$ symmetric TO modes, $A_1$ symmetric TO modes have larger nonlinear coefficients and smaller absorption coefficients. The four THz waves can be efficiently generated from COPG with a total photon conversion efficiency of 50.4%. The THz wave intensities can be enhanced by increasing the initial Stokes intensities.